\documentstyle[epsfig,preprint,aps]{revtex}

\begin{document}
\tightenlines

\newcommand{\mg}{\mbox{$^{26}$Mg}}
\newcommand{\al}{\mbox{$^{26}$Al}}
\newcommand{\alm}{\mbox{$^{26m}$Al}}
\newcommand{\alg}{\mbox{$^{26gs}$Al}}
\newcommand{\clg}{\mbox{$^{34gs}$Cl}}
\newcommand{\clm}{\mbox{$^{34m}$Cl}}
\newcommand{\cl}{\mbox{$^{34}$Cl}}
\newcommand{\msol}{\mbox{$M_\odot$}}
\newcommand{\zADNDT}{{\it Atomic Data and Nuclear Data Tables}}
\newcommand{\zmnras}{{MNRAS}}
\newcommand{\zaa}{{\it Astron. Astrophys.}}
\newcommand{\zapj}{{\it Astrophys. J.}}
\newcommand{\zapjs}{{\it Astrophys. J. Sup.}}
\newcommand{\zapjl}{{\it Astrophys. J. Letters}}
\newcommand{\znp}{{\it Nucl.~Phys.}}
\newcommand{\zpl}{{\it Phys.~Lett.}}
\newcommand{\zpr}{{\it Phys.~Rev.}}
\newcommand{\zprc}{{\it Phys.~Rev.~C}}
\newcommand{\znim}{{\it Nucl.~Inst.~and~Meth.}}
\def\power#1{\mbox{$\times10^{#1}\ $}}
\newcommand{\gap}{\mathrel{ \rlap{\raise.5ex\hbox{$>$}}
                      {\lower.5ex\hbox{$\sim$}}  } }
\newcommand{\lap}{\mathrel{ \rlap{\raise.5ex\hbox{$<$}}
		      {\lower.5ex\hbox{$\sim$}}  } }

\title{LIFETIMES OF \al\ AND \cl\ IN AN ASTROPHYSICAL PLASMA
}

\author{Alain Coc, Marie--Genevi\`eve Porquet}
\address{Centre de Spectrom\'etrie Nucl\'eaire et de Spectrom\'etrie
de Masse, IN2P3-CNRS and Universit\'e Paris Sud,\\ 
91405 Orsay Campus, France\\
}

\author{Fr\'ed\'eric Nowacki}
\address{Laboratoire de Physique Th\'eorique de Strasbourg,
3-5 rue de L'Universit\'e,\\
67084 Strasbourg Cedex 2, France}

\maketitle

\begin{abstract}
We study here the onset of thermal equilibrium affecting the lifetimes
of \al\ and \cl\ nuclei within a hot astrophysical photon gas. 
The \al\ isotope is of prime interest for gamma 
ray astronomy with the observation of its delayed 
($t_{1\over2}$=0.74~My) 1.809~MeV gamma--ray line. 
Its nucleosynthesis is complicated by the presence of a short lived 
($t_{1\over2}$=6.34~s) spin isomer.
A similar configuration is found in \cl\ where the decay of its 
isomer ($^{34m}$Cl, $t_{1\over2}$=32~m) 
is followed by delayed gamma--ray emission with
characteristic energies.
The lifetimes of such nuclei are reduced at high temperature by the 
thermal population of shorter lived levels. 
However, thermal equilibrium within \al\ and \cl\ levels is delayed 
by the presence of the isomer.
We study here the transition to thermal equilibrium where branching 
ratios
for radiative transitions are needed in order to calculate lifetimes.
Since some of these very small branching ratios are not known 
experimentally, we use results of shell model calculations. 
\end{abstract}



\section{Introduction}

The discovery and the subsequent mapping of \al\ in the interstellar
medium through the detection of its 1809 keV $\gamma$-ray line
by satellites (HEAO-3, COMPTEL-CGRO, and for the future: INTEGRAL)
has increased the interest in \al\ nucleosynthesis.
The potential sites for \al\ production include supernovae, 
Wolf-Rayet stars, AGB (Asymptotic Giant Branch) stars and novae 
(see ref.~\cite{Pra96} for a review).
Consequently, the nuclear physics involved in \al\ nucleosynthesis is
of renewed importance.
Thermonuclear reaction rates involved in \al\ production and 
destruction are discussed in a recent compilation\cite{NACRE}.
Here we discuss its off-equilibrium
destruction rate through the thermal population of excited levels and
in particular of its isomer.
If beta decay probabilities are available, lifetimes at 
thermal equilibrium can be readily calculated.
However, thermal equilibrium between the ground and isomeric
states are delayed by the large spin difference.
This was first studied by Ward and Fowler\cite{WF80} who set the rule
used in \al\ nucleosynthesis calculations, stating that \alg\ and 
\alm\ have to be considered either as separate isotopes (no 
equilibrium) at temperatures below $\approx$ 0.4~GK or as a single 
one (full equilibrium) above.
This temperature is uncomfortably close to the peak temperature 
attained in nova outbursts ($\lap$0.35~GK) or in AGB thermal pulses 
($\lap$0.3~GK). Moreover, some of the nuclear data used by 
Ward and Fowler are estimates that are known to be valid only to 
within many orders of magnitude.
Hence, we felt that it was time to reconsider the {\it onset} of 
thermal equilibrium in \al. The \al\ effective lifetime has been 
calculated\cite{Tours} using systematics of radiative transition 
probabilities to evaluate unknown transitions in  \al. 
Here we use instead results of shell model calculations.

The origin for the presence of an isomer in \al\ is well known. 
It is an odd--odd nucleus with $N=Z$ (=13). Hence, the two unpaired
nucleons are in the same $nlj$ shell ($j={5\over2}$).
The most favored configurations are when they couple to $J^\pi$
= $0^+$ or $2j^+$ (5$^+$) making the ground state (5$^+$) and isomer 
(0$^+$).
Internal (gamma) transitions from the isomer to the ground state are
inhibited by the large spin difference.
Beta decay of the isomer ($0^+$) to the ground state of the
even-even nucleus $^{26}$Mg ($0^+$) is not hampered by spin difference
and hence constitutes the sole decay mode of \alm\ 
($t_{1\over2}$=6.34~s).
On the contrary the ground state \alg\ with $J^\pi=5^+$ has its beta 
decay to the $^{26}$Mg ground state forbidden. Hence it decays slowly
($t_{1\over2}$=0.74~My) towards excited sates of  $^{26}$Mg which
subsequently de--excite through internal transitions leading to the 
observed gammas.

A similar configuration occurs in \cl\ save that the unpaired nucleons
belong to a $j={3\over2}$ shell, making the situation less contrasted.
It has also $Z=N$ (=17), odd--odd, and a relatively long lived isomer.
However, contrary to the \al\ case, the ground state has the lowest 
spin and hence a shorter half-life ($0^+$, 1.53~s) than the isomer
($3^+$, 32~m), located 146~keV above.
It is also \clm\ which decays by beta emission towards excited states 
of $^{34}$S, and thus emits delayed gammas while \clg\ decays to the
$^{34}$S ground state.
So on a much shorter time scale, \cl\ has properties similar to \al\
when considering the two first levels with high (3$^+$;5$^+$) and low 
(0$^+$) spins save that there is an inversion in their relative 
location.
The lower spin of \clm\ compared to \alm\ results in a shorter 
lifetime with respect to beta decay and to a significant decay 
through internal transition to \clg.

Due to its short half life, gamma ray lines resulting from \clm\ 
decay could only
be seen in events where the ambient medium becomes transparent
to gamma rays in a matter of hours after the nucleosynthesis phase.
This happens in nova outbursts where \cl\ has been considered
as a source of 511~keV gamma rays\cite{Lei87}.
However, \clm\ delayed gamma emission includes also specific lines
($E_\gamma$ = 2.128 and 3.304~MeV, with 43\% and 12\% branching
ratio respectively\cite{Endt90} not considered in ref.~\cite{Lei87}) 
which are observationally more interesting.
Synthetic gamma ray spectra of novae have been calculated\cite{Gom98,Her99}
including gamma ray emission following the decay of $^{18}$F
whose half--life (110~m) is of the same order of magnitude as for 
\clm\ (32~m). According to these calculations, gamma rays following 
$^{18}$F decay could be detected a few hours after the outburst. 
In consequence \clm\ should be considered as a potential source of 
observable gamma ray lines but with a prerequisite study of its modes
of production and destruction and in particular its lifetime under 
astrophysical conditions.

For \al\ and \cl\ internal transition probabilities, we use 
experimental data when available or results of shell model 
calculations otherwise.
To check the reliability of these calculations, we first compare 
theoretical and experimental radiative widths in \al, \cl\ and 
neighboring nuclei and deduce confidence limits for the shell model 
results.
Then we calculate the transition to equilibrium with these new values.
For this purpose, we consider the first levels as separate nuclides 
and all possible internal transitions between them together with 
their beta decays.
The set of equations is solved numerically and the results are 
provided as an analytical fit.

\section{Lifetime of nuclei in a plasma}

At high densities, lifetimes of nuclei are reduced due to the 
increasing Fermi energy of the electrons that opens up electron 
capture channels otherwise energetically forbidden.
This effect occurs at densities above $\rho\approx10^5$\cite{FFN80}
when the Fermi energy of the electrons,
[$U_F=m_e(((3\pi^2)^{2/3}(\hbar/m_e)^2N_A^{2/3}
(Z\rho/A)^{2\over3}+1)^{1\over2}-1)$] 
reaches a few tens keV.
Here we concentrate on the onset of equilibrium at moderate densities
($\rho\lap10^5$) that prevail, for instance, in nova outburst.
The principles of the calculation are presented in ref.~\cite{WF80} 
but it is worth while to give here a short summary, taking \al\ as a 
typical example.


Equilibration between \alm\ and \alg\, proceeds through
intermediate $E_x\lap$1~MeV levels\cite{WF80}. 
The corresponding two levels (labeled a and b) together with the 
gamma ray transitions that link them are displayed in 
Fig.~\ref{f:al26sc}.
The thick or thin arrows correspond to transition
probabilities ($\lambda_{ij}$) that are known experimentally or, 
respectively, have to be estimated. 
They are labeled by their electric ($EL$) or magnetic ($ML$) 
multipolarity ($L$).
No arrow links \alm\ and \alg\ since a transition with such a high 
multipolarity ($M5$) is strongly inhibited 
even though this rate is slightly enhanced by the photon 
bath\cite{WF80}.
These two levels are only connected indirectly through transitions 
via the higher lying levels and in particular those located at 
E$_{\mathrm cm}$=0.417 and 1.058~MeV.

Within stellar environment, the gamma transition probabilities are
modified by the thermal photon gas and we use the method
exposed in ref.~\cite{WF80}.
Let $i$ and $j$ two levels such that $E_j>E_i$ and where the only 
mode of deexcitation for level $j$ is a gamma transition to $i$.
The evolution of the populations ($N_j$, $N_i$) of these levels is
governed by the set of coupled equations\cite{Atom}:

\begin{eqnarray*}
{{dN_j}\over{dt}} & = & 
-\lambda_{ij}N_j+(-\lambda_{ij}N_j+\lambda_{ji}N_i)u(T)\\
{{dN_i}\over{dt}}& =& 
\lambda_{ij}N_j+(\lambda_{ij}N_j-\lambda_{ji}N_i)u(T)\\
\end{eqnarray*}

The first term of the second member represents the spontaneous decay
($j\to i$), the following the stimulated ($j\to i$) and induced
($i\to j$) transitions and $u(T)$ is the photon density:

\begin{eqnarray*}
u(T) & = &\left(\exp\left({{E_j-E_i}\over{kT}}\right)-1\right)^{-1}\\
\end{eqnarray*}

The $\lambda$ coefficients are readily obtained by considering the 
limits $i$) $T\to0$ where only spontaneous decay occurs, 
and $ii$) thermal equilibrium (${{dN_i}\over{dt}}=0$), that is :

\begin{eqnarray*}
i)\;\;\hbar\lambda_{ij} &= &\Gamma_{\gamma;j} \\
ii)\;\;{{\lambda_{ji}}\over{\lambda_{ij}}} & = &
{{2J_j+1}\over{2J_i+1}}\exp\left({{E_j-E_i}\over{kT}}\right)\\
\end{eqnarray*}

To be more general, the evolution of the population of the
various levels ($i,j\in\;\{o,m,a,b,..\}$) linked by all possible 
internal gamma transition is represented by a set of linear 
differential equations which can be written in matrix form as:
$d{\mathbf N}/dt = -{\mathbf\lambda}{\mathbf  N}$.
This set of equations can be readily solved numerically by using a 
standard implicit code for nucleosynthesis\cite{Tours} using the 
Arnett and Truran\cite{Arn69} prescription.
For this purpose, the relevant nuclear levels ($o,m,a,b,..$) are 
introduced as separate ``isotopes'' connected via ``nuclear 
reactions'' (i.e. gamma transitions) whose rates are given by the 
$\lambda_{ij}$ matrix elements. 
To these ``reactions'', one must add beta decay rates from the
various levels to the daughter nucleus. When not available 
experimentally (i.e. for levels above the ground and isomeric states) 
these beta decay rates are obtained from shell model 
calculations\cite{Kaj88}. 
(At $\rho\gap10^5$, one must take into account that $\log(ft)$ 
depends on the electronic density and hence on $\rho$~\cite{FFN80}.
This is not considered here where we limit ourselves to {\it low} 
densities.)
Then, starting with initial abundance such that only the ground or 
isomeric state are populated one can obtain their lifetime by 
calculating numerically the time needed for the initial abundance 
to be reduced by a $1/e$ factor.

\section{Available nuclear data}

In \al, the known transitions linking the $3^+$ level with the ground 
state and the $1^+$ level with the isomer, are an  $E2$, and a $M1$,
respectively (Fig.~\ref{f:al26sc}).
However, the possibility remains of a $M3$ transition linking the
$3^+$ and isomeric levels and of an $E2$ between the $1^+$ and
$3^+$ levels\cite{WF80}.
Even though the branching ratios $M3/E2$ and $E2/M1$ are
expected to be very small, they provide the links between the ground
state and isomer but delay the onset of equilibrium.
These branching ratios are too small to be measured and accordingly,
Ward and Fowler\cite{WF80} had to estimate them.
They assumed one Weisskopf {\it unit} (W.u.) for electric transitions
($EL$) or one Moszkowski {\it unit} (M.u.) for magnetic ones ($ML$).
However, the value of 1.~M.u/1.~W.u. they used for the
$M3(a{\to}m)$/$E2(a{\to}o)$ lies well outside the range 
(7 orders of magnitude\cite{Tours,Rou98}) spanned by experimental 
values.
Indeed, the $a{\to}o$ transition has been measured and corresponds to 
7.7~W.u., this is equivalent to assume 7.7~M.u. for the $a{\to}m$ 
transition which is above the upper limit $\approx$1~M.u. derived 
from statistics of experimental values of $M3$ transition 
probabilities\cite{Rou98}.
For the other branching ratio ($E2/M1$), the choice of 1~W.u./1~M.u.
is compatible with the $M1$ experimental value of 2.4~M.u. and the 
wide range of $E2$ reduced transition strengths\cite{Endt79}
($\approx10^{-2}$--10$^{2}$~W.u).


In \cl, contrary to the \al\ case, an internal ($M3$, $m{\to}o$) 
transition links the isomer to the ground state with a 44.6\% 
branching ratio (Fig.~\ref{f:cl34sc}). 
The two $1^+$ levels ($a$ and $b$ at 0.461 and 0.666~MeV) decay 
to the ground state with known probabilities\cite{ToI,Endt90}.
Only upper limits for the branching ratios are available for the
transition to the isomer ($M1$, $a{\to}m$ and $b{\to}m$) or between
the two $1^+$ levels ($M1+E2$, $b{\to}a$).
The next level ($2^+$) is located at 1.230~MeV and has known radiative
transition probabilities to all the \cl\ levels below save for the 
ground state where only an upper limit is available. 
These upper limits are well above those provided by statistics 
of radiative reduced transition strengths\cite{Endt79}. 
They only reflect experimental limitations and accordingly have not 
been considered in our calculations. As for the \al\ case, 
experimentally known transitions are represented by thick arrows 
(Fig.~\ref{f:cl34sc}) while thin arrows represent missing data that 
have to be obtained theoretically. 

\section{Shell Model calculations}

Instead of using {\it one} Weisskopf or Moszkowski {\it unit} as 
estimates of radiative transition probabilities, we use shell model 
calculations to provide the unknown ones. We also compare calculated 
and experimental values in the neighboring nuclei in order to 
estimate confidence limits for the shell model results.


The  calculations include the full sd shell 
(d$_{\frac{5}{2}}$,s$_{\frac{1}{2}}$
and d$_{\frac{3}{2}}$ orbitals) for protons and neutrons.
Here we use the USD interaction of Wildenthal \cite{usd}.
We have also done calculations with the two sets of the 
Chung-Wildenthal (CW) \cite{cw} interaction (lower and upper sd part)
in order to compare with USD results.
Indeed, the USD interaction is designed for the full A=16-40 mass
range and therefore should provide the most reliable values but 
comparison with CW calculations allows to appreciate the 
uncertainties involved in such calculations.
The diagonalization procedure and  transition calculations
are performed using the shell model code Antoine \cite{ant1,ant2}.
For the calculations of electromagnetic transitions, we have used
a polarization effective charge of 0.5, and the following gyromagnetic
factors, $g_s(p)=$-5.59, $g_l(p)=$1.0, $g_s(n)=$3.83 and $g_l(n)=$0.

All the transition probabilities calculated for \al, its odd-A 
neighbors ($^{25}$Al and $^{25}$Mg), \cl\ and  its odd-A neighbours 
($^{33}$Cl and $^{33}$S) are given in Tables I to III in comparison
with the experimental values.
Fig.~\ref{f:expth} shows that most of the calculated radiative widths
lie within a factor of three from the experimental values. 
Accordingly, the upper and lower limits of the transition 
probabilities used in the present calculations are obtained by 
multiplying the theoretical values by a factor of 3 and 1/3 
respectively.

\section{Results for \al\ and \cl\ effective lifetimes}

To calculate \al\ and \cl\ lifetimes we use experimental or 
theoretical data (Table I, II and III) as discussed in the preceding
sections following the method exposed in the second section. 
We emphasize that by numerically
solving the set of coupled differential equations representing all the
transitions ($\gamma$ and $\beta$) displayed in Fig.~\ref{f:al26sc} and
Fig.~\ref{f:cl34sc} we make no assumption, at any stage, on the degree
of equilibrium achieved. Accordingly, the main decay channel 
(e.g. internal transition or $\beta$ decay from the $a$ level in \al)
is the result of the calculation.

\subsection{\al}

With this method, Coc and Porquet\cite{Tours} have already compared
the two formulas\cite{WF80,Vog89} giving the \alg\ off--equilibrium 
effective lifetime. 
Ward and Fowler\cite{WF80} assumed that in these condition \al\
decays to \mg\ predominantly through the beta decay of the $3^+$ (a) 
level while Vogelaar\cite{Vog89} favors the decay from the $0^+$ (m) 
level.
When the calculations are made numerically, without {\it a priori}
assumptions, the only significant decay channel\cite{Tours} is the
one assumed by Vogelaar\cite{Vog89} i.e. through internal transition 
to the isomer followed by its beta decay. Only in the extreme case, 
where the $M3(a{\to}m)$ transition probability is assumed to be 
around $10^{-3}$~M.u.
does beta decay from the $3^+$ level become significant as the 
transition to the $0^+$ level is strongly inhibited. 
This very small value corresponds to the lower limit deduced from 
the statistics of $M3$ reduced strength\cite{Rou98} but can be 
rejected on the basis of shell model calculations ($\approx$1~M.u.).

Using now the updated nuclear data, the \al\ effective life time is 
depicted in Fig.~\ref{f:alife} by a solid line.
(For comparison, the dash-dotted line represents the result of the 
Ward and Fowler\cite{WF80} off--equilibrium formula.) 
The hatched area represents the uncertainty due to the confidence 
limits assigned to the shell model calculations as discussed above. 
The resulting uncertainty is too small to have any
consequence in astrophysics. 
Below $\approx$0.15~GK, the lifetime is equal
to its laboratory value. Above $\approx$0.4~GK it is equal to its
equilibrium value $\lambda_{eq} =  9.9\times10^{-3}\;
\exp\left({{-2.651}\over{T_9}}\right)$\cite{WF80} represented 
by a dashed curve. 
Below, it can be approximated (dashed curve) by using the following 
formula which gives the transition probability (i.e. the inverse of 
the effective lifetime)

\begin{eqnarray*}
\lambda_{off} & = & 2.97\times10^{-14} + 
4.07\times10^{-2}\;\exp\left({{-4.839}\over{T_9}}\right)+ \\
&  &2.10\times10^{8}\;\exp\left({{-12.28}\over{T_9}}\right)\\
\end{eqnarray*}


Here, as usual in astrophysics, $T_9$ represents the temperature in 
units of GK ($10^9$~K).
The first term corresponds to the \alg\ laboratory decay. 
The two last terms originate from the population of the 3$^+$ level,
$\left({{2J_o+1}\over{2J_a+1}}
\exp\left(-{{E_a}\over{KT}}\right)\right)$
times the probability that it decays to the isomeric level 
($\lambda_{ma}$) or that it undergoes an induced transition to 
the 1$^+$ level 
$\left(\lambda_{ba}\equiv\lambda_{ab}{{2J_b+1}\over{2J_a+1}}
\exp\left(-{{E_b-E_a}\over{KT}}\right)\right)$.
As shown in Fig.~\ref{f:alife}, the two dashed curves representing the
equilibrium and off--equilibrium lifetimes meet around 0.4~GK. 
Hence, as suggested in ref.~\cite{Tours}, the temperature below which
\alg\ and \alm\ have to be considered as separate nucleides is not 
affected by the use of new data.

\subsection{\cl}

Figure~\ref{f:clife} displays the effective \cl\ lifetime as a 
function of temperature with the same convention as for \al\ in 
Fig.~\ref{f:alife}.
The transition between the laboratory lifetime to the equilibrium 
lifetime occurs between 0.12 and 0.25~GK and the remaining nuclear 
uncertainties have a negigible effect (dashed area). 
The decay rate at equilibrium is given by:

\begin{eqnarray*}
\lambda_{eq}& = &
{{0.454 + 1.40\times10^{-3}\exp\left({{-1.699}\over{T_9}}\right)}
\over{1 + 7\exp\left({{-1.699}\over{T_9}}\right)}}
\end{eqnarray*}

Contrary to the \al\ case, at equilibrium, the lifetime increases with
temperature following the thermal depopulation of the short lived 
ground state. At lower temperature, \clm\ and \clg\ have to be 
considered separately and the \clm\ decay rate is approximated by:

\begin{eqnarray*}
\lambda_{off}& = & 3.61\times10^{-4} + 
8.77\times10^{6}\exp\left({{-3.651}\over{T_9}}\right)\\
\end{eqnarray*}


The results obtained with these formula are represented by dashed 
lines on Fig.~\ref{f:clife}. They meet at $T\approx$0.22~GK which 
marks the limit for equilibrium. This last formula approximates 
very well the result of the numerical calculation. 
Since it includes only the effect of the $m{\to}a$ transition, 
it shows that only this transition has a significant influence 
on \clm\ lifetime.

\section{Conclusions}

We calculated by numerical integration of the coupled differential 
equations the lifetimes of \al\ and \cl\ in a hot astrophysical 
photon gas assuming {\it low} densities ($\lap10^5$~g/cm$^3$). 
We obtained crucial radiative transition probabilities, not available 
experimentally, from  shell model calculations.
The temperature (0.4~GK) which limits the domain where \alg\ and \alm\
have to be considered separately is insensitive to the remaining 
nuclear uncertainties.
Below this temperature and down to $\approx$0.16~GK, the \alg\ 
lifetime is shorter (by up to four orders of magnitude) than the one
proposed by Ward and Fowler. The discrepancy comes from their 
hypothesis on the main \al\ decay channel (beta decay from the 3$^+$
level) which is not correct as pointed out by Vogelaar.    
Below 0.22~GK, \clg\ and \clm\ are not at equilibrium and have to be 
treated as two separate nuclides. The \cl\ lifetime drops rapidly 
above 0.15~GK (two orders of magitudes between 0.15 and 0.2~GK).   
We provide analytical formulas that approximate the \alg\ and \clm\
lifetimes when they are not yet in equilibrium with \alm\ and \clg\
respectively. 

Considering the various \al\ potential astrophysical sources (i.e.
temperature and density conditions), it is not clear whether this 
modified lifetime will affect significantly its production.
On the contrary \cl\ is only of interest for novae and only the 
hottest ones can synthetize isotopes in the S--Ar region. 
Unfortunately, our calculations show that in the conditions that 
prevail in such events (up to $\approx$0.35~GK), \clm\ is efficiently
destroyed by induced transition to the short--lived \clg.

\section{Acknowledgments}

We are grateful to E.~Caurier, H.T.~Duong, B.~Roussi\`ere and 
D.~Lunney for valuable discussions. 
This work was partially supported by PICS 319. 


\begin{figure}[h] 
\caption{Level scheme for $^{26}$Al. The first levels of $^{26}$Al 
with their characteristics are represented together with the beta and 
gamma transitions considered in the calculations. The thick arrows 
represent the experimentally known transitions while the thin ones 
come from shell model calculations. 
The matrix element $\lambda_{ij}$ represents the
transition probability from level $j$ to level $i$.}
\label{f:al26sc}
\end{figure}

\begin{figure} 
\caption{Level scheme for \cl. Conventions are the same as Fig.~1 
(see Tables for details.)}
\label{f:cl34sc}
\end{figure}

\begin{figure}[h] 
\caption{Comparison between experimental and theoretical gamma widths.
Most calculated widths deviate from experimental data by less than a 
factor of three (dashed lines.)}
\label{f:expth}
\end{figure}

\begin{figure}[h] 
\caption{
The effective lifetime of \al\ as a function of temperature
calculated using shell model transition probabilities (solid line).
The hatched area shows the corresponding uncertainty (see text.)
The analytic formulas for $\lambda_{off}$ and $\lambda_{eq}$  
reproduce the effective lifetime (dashed curves) on both sides of 
the equilibrium temperature.
The dash-dotted curve corresponds to the formula provided by Ward and
Fowler\protect\cite{WF80}}
\label{f:alife}
\end{figure}

\begin{figure}
\caption{Effective lifetime of \clm\ as a function of temperature. 
(Same conventions as in Fig.~4.)}
\label{f:clife}
\end{figure}

\onecolumn

\widetext
\begin{table*}[h]
\caption[]{Comparison of M3 transitions}
\begin{tabular}{|c|c|c|c|c|c|c|}
\hline\noalign{\smallskip}
\hline\noalign{\smallskip}
Nuclei & Transition  & $t_{1\over2}$& $E_\gamma$ & 
$\gamma$--branching & $B(M3)$ & $B(M3)$\\
&&&& ratio & Exp. & USD \\
&&&(MeV)  & & ($\mu_N^2\;fm^4$)& ($\mu_N^2\;fm^4$)\\
\hline\noalign{\smallskip}
$^{24}$Na & $0.472;1^+\;\to\;0;4^+$
& 20.2$\pm$0.07 ms & 0.472& 0.9995 & 1046$\pm$4& 1795.\\
\hline\noalign{\smallskip}
$^{24}$Al & $0.426;1^+\;\to\;0;4^+$
&131.3$\pm$2.5 ms & 0.426&  0.82 & 270$\pm$12& 725. \\
\hline\noalign{\smallskip}
$^{26}$Al & $0.417;3^+\;\to\;0.228;0^+$
& 1.25$\pm$0.03 ns &0.189 & unknown & -- & 1206.\\
\hline\noalign{\smallskip}
$^{34}$Cl & $0.146;3^+\;\to\;0;0^+$
& 32.00$\pm$0.04 m&0.146  & 0.381 &15.2$\pm$0.2& 18.0\\
$^{38}$Cl &$0.671;5^-\;\to\;0;2^-$ 
& 715$\pm$3 ms &0.671  &1. &2.52$\pm$0.01& 0.008\\
\hline\noalign{\smallskip}
\noalign{\smallskip}\hline
\end{tabular}
\label{t:bm3}
\end{table*}
\noindent

\widetext
\begin{table*}[h]
\caption[]{Comparison of E2 transitions}
\begin{tabular}{|c|c|c|c|c|c|}
\hline\noalign{\smallskip}
\hline\noalign{\smallskip}
Nuclei & Transition & $t_{1\over2}$& $E_\gamma$& $B(E2)$& $B(E2)$\\
&&&& Exp. & USD \\
&& &(MeV)  & ($e^2\;fm^4$) &($e^2\;fm^4$)  \\
\hline\noalign{\smallskip}
\hline\noalign{\smallskip}
$^{25}$Mg & $0.585;1/2^+\;\to\;0;5/2^+$
& 3.38$\pm$0.05 ns& 0.585 &$2.44\pm0.04$& 33.3\\
$^{25}$Mg & $0.975;3/2^+\;\to\;0;5/2^+$
& 11.3$\pm$0.3 ps& 0.975 & $3.5\pm0.3$& 4.3\\
$^{25}$Mg & $0.975;3/2^+\;\to\;0.585;1/2^+$
&  11.3$\pm$0.3 ps & 0.390&$49\pm22$& 65\\
\hline\noalign{\smallskip}
$^{25}$Al & $0.452;1/2^+\;\to\;0;5/2^+$
& 2.29$\pm$0.03 ns& 0.452 &$13.2\pm0.2$& 14.6\\
$^{25}$Al & $0.945;3/2^+\;\to\;0;5/2^+$
& 4.3$\pm$1.1 ps& 0.945 & $8\pm3$& 8.6\\
$^{25}$Al & $0.945;3/2^+\;\to\;0.452;1/2^+$
& 4.3$\pm$1.1 ps& 0.493 & $10\pm10$& 74.6\\
\hline\noalign{\smallskip}
$^{26}$Al & $0.417;3^+\;\to\;0;5^+$
& 1.25$\pm$0.3 ns& 0.417 & 36$\pm$1& 48.\\
$^{26}$Al & $1.058;1^+\;\to\;0.417;3^+$
& 25$\pm$5 fs& 0.641 & unknown & 5.8\\
\hline\noalign{\smallskip}
$^{33}$Cl & $1.986;5/2^+\;\to\;0;3/2^+$
& 55$\pm$11 fs& 1.986 &49$\pm22$& 65\\
$^{33}$Cl & $1.986;5/2^+\;\to\;0.811;3/2^+$
& 55$\pm$11 fs & 1.176&$<$133& 25\\
\hline\noalign{\smallskip}
$^{33}$S & $0.841;1/2^+\;\to\;0;3/2^+$
& 1.17$\pm$0.03 ps & 0.841&26.4$\pm1.5$& 20\\
$^{33}$S & $1.967;5/2^+\;\to\;0;3/2^+$
& 104$\pm$14 fs& 1.967 &44$\pm7$& 55\\
$^{33}$S & $1.967;5/2^+\;\to\;0.841;1/2^+$
& 104$\pm$14 fs& 1.126 &39$\pm10$& 19\\
\hline\noalign{\smallskip}
$^{34}$Cl & $0.461;1^+\;\to\;0.146;3^+$
& 5.2$\pm$0.3 ps & 0.315&$<$178.& 5.44\\
$^{34}$Cl & $0.666;1^+\;\to\;0.461;1^+$
& 9.1$\pm$0.6 ps & 0.205&$<$1724.& 20.8\\
$^{34}$Cl & $0.666;1^+\;\to\;0.146;3^+$
& 9.1$\pm$0.6 ps & 0.519&$<$16.6& 18.5\\
$^{34}$Cl & $1.230;2^+\;\to\;0.661;1^+$
& 13.7$\pm$0.9 ps& 0.565 &33$\pm$11& 24.9\\
$^{34}$Cl & $1.230;2^+\;\to\;0.461;1^+$
& 13.7$\pm$0.9 ps & 0.769&32$\pm$11& 25.7\\
$^{34}$Cl & $1.230;2^+\;\to\;0.146;3^+$
& 13.7$\pm$0.9 ps& 1.084 &4.2$\pm$3.1& 0.621\\
$^{34}$Cl & $1.230;2^+\;\to\;0;0^+$
& 13.7$\pm$0.9 ps & 1.230&$<$0.074& 0.221\\
\hline\noalign{\smallskip}
\noalign{\smallskip}\hline
\end{tabular}
\label{t:be2}
\end{table*}
\noindent

\widetext
\begin{table*}[h]
\caption[]{Comparison of M1 transitions}
\begin{tabular}{|c|c|c|c|c|c|}
\hline\noalign{\smallskip}
\hline\noalign{\smallskip}
Nuclei & Transition & $t_{1\over2}$ & $E_\gamma$ & 
$B(M1)$& $B(M1)$\\
&&&& Exp. & USD\\
&&&(MeV)   & ($\mu_N^2$)& ($\mu_N^2$) \\
\hline\noalign{\smallskip}
\hline\noalign{\smallskip}
$^{25}$Mg & $0.975;3/2^+\;\to\;0;5/2^+$
& 11.3$\pm$0.3 ps & 0.975&$(2.8\pm0.1)\times10^{-2}$& 0.035\\
$^{25}$Mg & $0.975;3/2^+\;\to\;0.585;1/2^+$
& 11.3$\pm$0.3 ps & 0.390&$(1.69\pm0.06)\times10^{-3}$& 0.008\\
\hline\noalign{\smallskip}
$^{25}$Al & $0.945;3/2^+\;\to\;0;5/2^+$
& 4.3$\pm$1.1 ps & 0.945& $(4.3\pm1.2)\times10^{-3}$& 0.13\\
$^{25}$Al & $0.945;3/2^+\;\to\;0.452;1/2^+$
& 4.3$\pm$1.1 ps& 0.493 & $(4.2\pm1.1)\times10^{-2}$& 0.066\\
\hline\noalign{\smallskip}
$^{26}$Al & $1.058;1^+\;\to\;0.228;0^+$
& 25$\pm$5 fs & 0.829& 2.76$\pm$0.55& 9.7\\
\hline\noalign{\smallskip}
$^{33}$Cl & $0.811;1/2^+\;\to\;0;3/2^+$
& 1.2$\pm$0.2 ps & 0.811&(6.2$\pm1.0)\times10^{-2}$& 0.063\\
$^{33}$Cl & $1.986;5/2^+\;\to\;0;3/2^+$
& 55$\pm$11 fs& 1.986 &(7.6$\pm1.6)\times10^{-2}$& 0.017\\
\hline\noalign{\smallskip}
$^{33}$S & $0.841;1/2^+\;\to\;0;3/2^+$
& 1.17$\pm$0.03 ps& 0.841 &(5.5$\pm0.2)\times10^{-2}$& 0.031\\
$^{33}$S & $1.967;5/2^+\;\to\;0;3/2^+$
& 104$\pm$14 fs & 1.967&(3.7$\pm0.5)\times10^{-2}$& 0.008\\
\hline\noalign{\smallskip}
$^{34}$Cl & $1.230;2^+\;\to\;0.666;1^+$
& 13.7$\pm$0.9 ps & 0.565&(5.0$\pm0.5)\times10^{-3}$& 0.0072\\
$^{34}$Cl & $1.230;2^+\;\to\;0.461;1^+$
& 13.7$\pm$0.9 ps & 0.769&$(8.8\pm4.5)\times10^{-4}$& 0.0025\\
$^{34}$Cl & $1.230;2^+\;\to\;0.146;3^+$
& 13.7$\pm$0.9 ps & 1.084&$(3.2\pm2.5)\times10^{-4}$& 0.0016\\
$^{34}$Cl & $0.666;1^+\;\to\;0.461;1^+$
& 9.1$\pm$0.6 ps& 0.205 & $<$0.005& 0.0007\\
$^{34}$Cl & $0.666;1^+\;\to\;0;0^+$
& 9.1$\pm$0.6 ps & 0.666& $(1.47\pm0.10)\times10^{-2}$& 0.018\\
$^{34}$Cl & $0.461;1^+\;\to\;0;0^+$
& 5.2$\pm$0.3 ps& 0.461 & $(7.74\pm4.6)\times10^{-2}$& 0.110\\
\hline\noalign{\smallskip}
\noalign{\smallskip}\hline
\end{tabular}
\label{t:bm1}
\end{table*}
\noindent

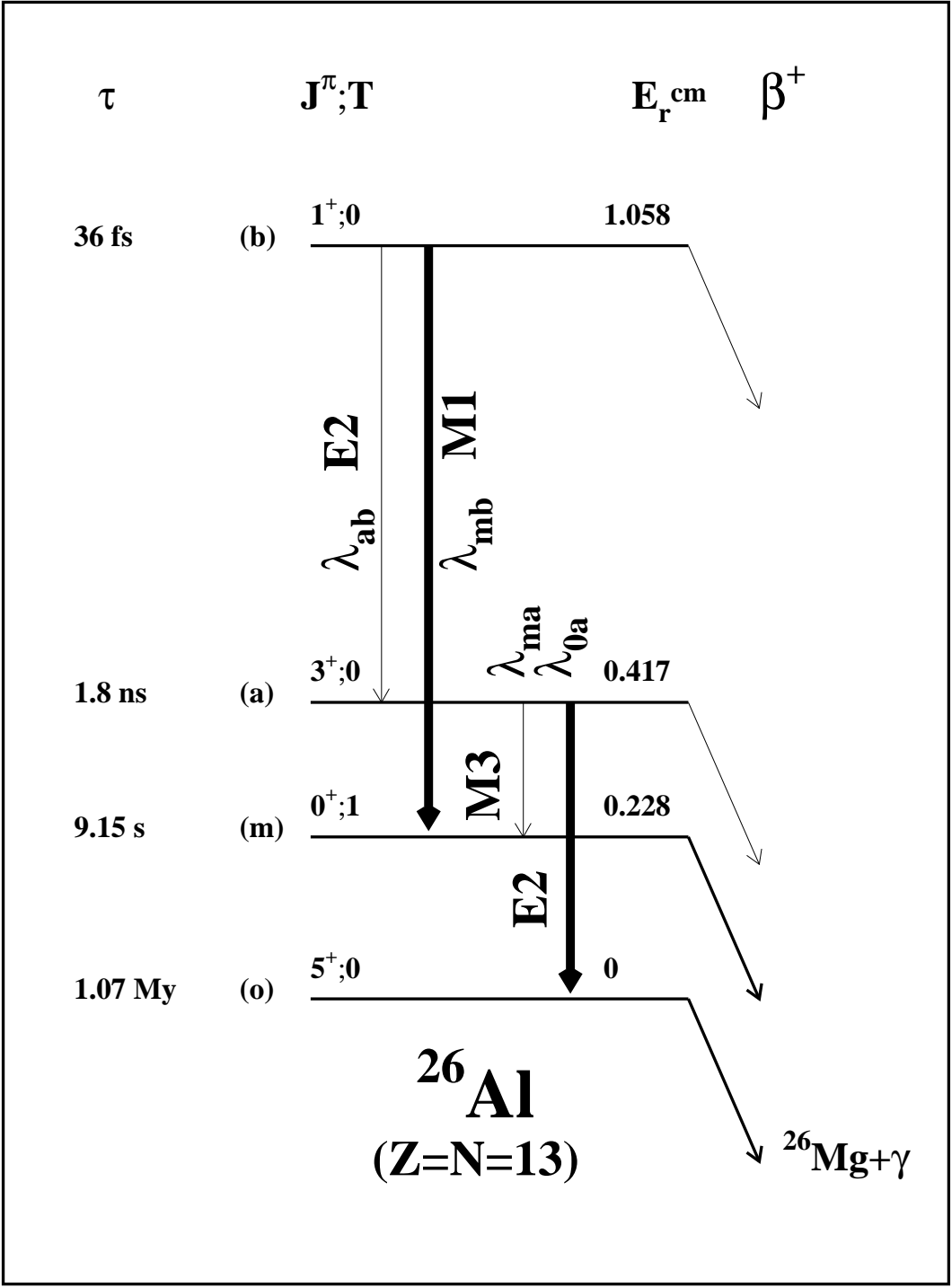
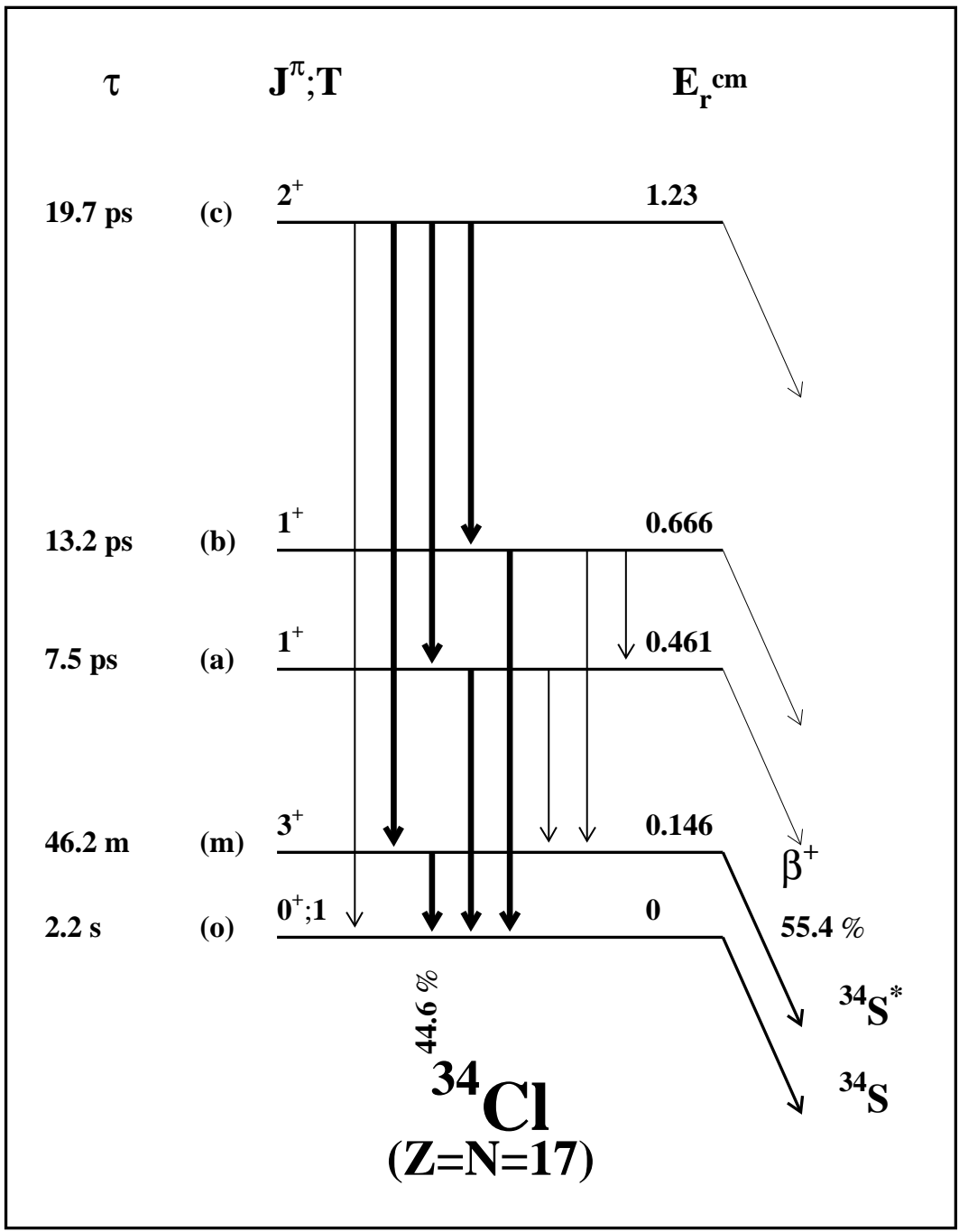
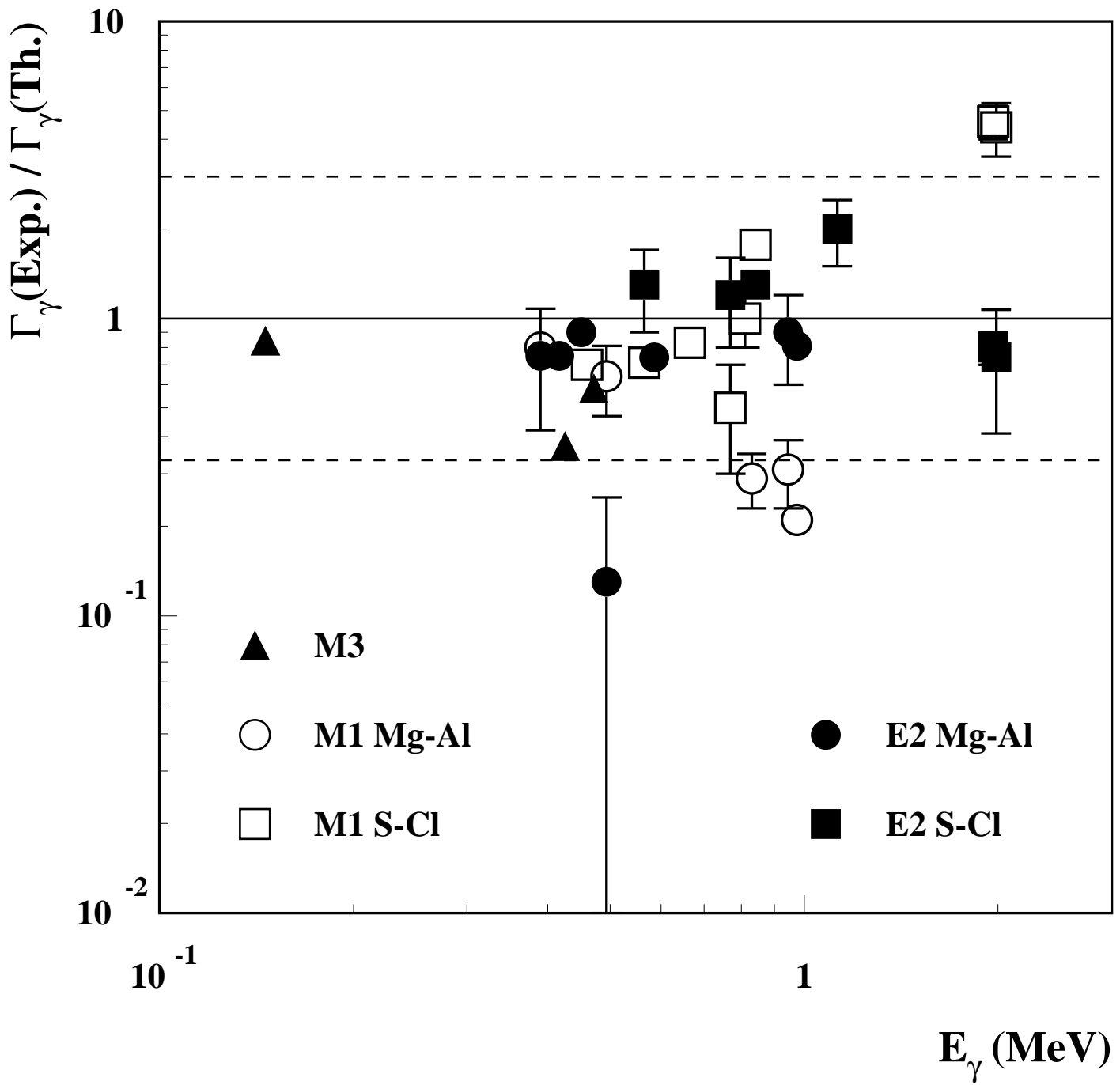
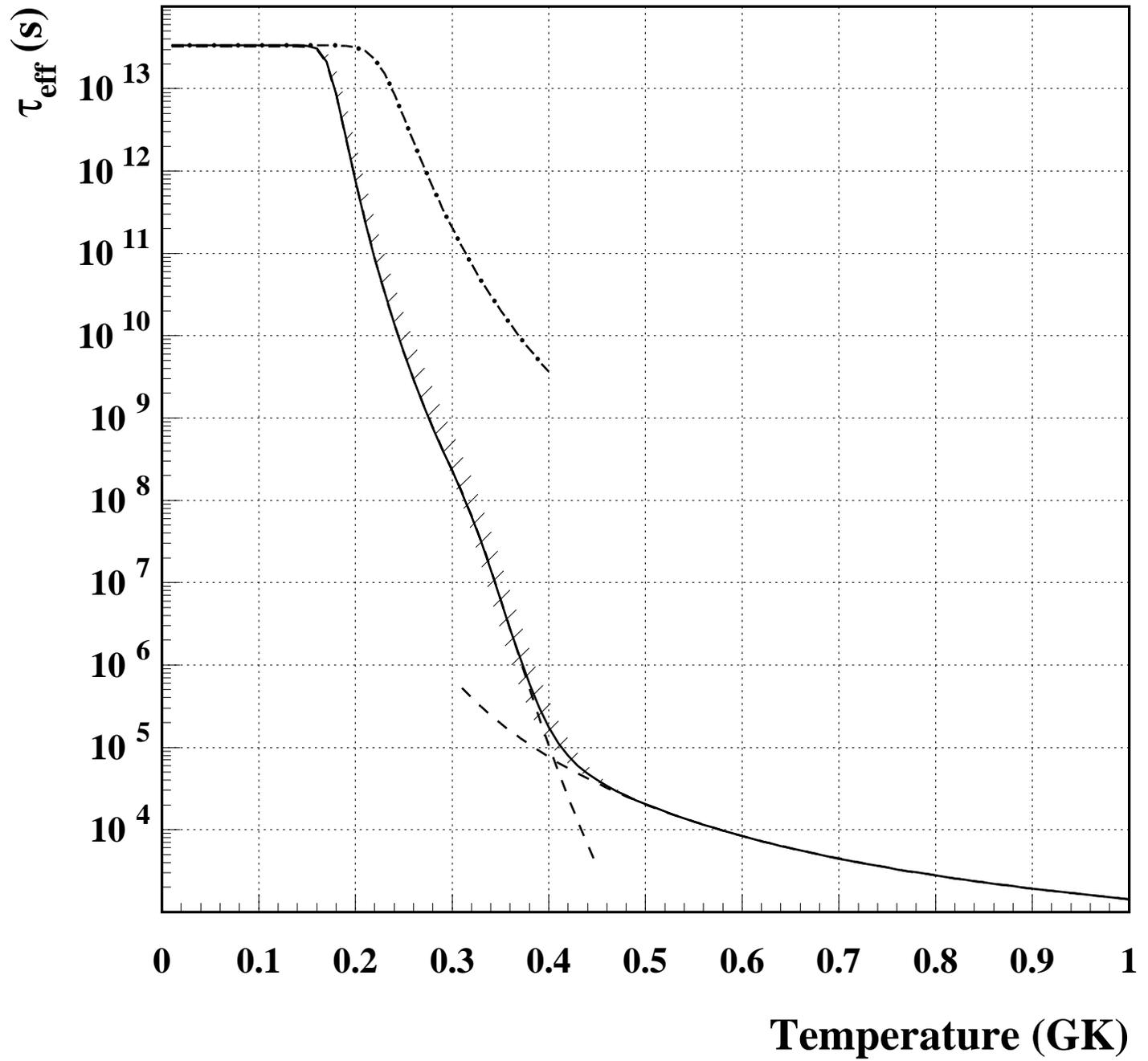
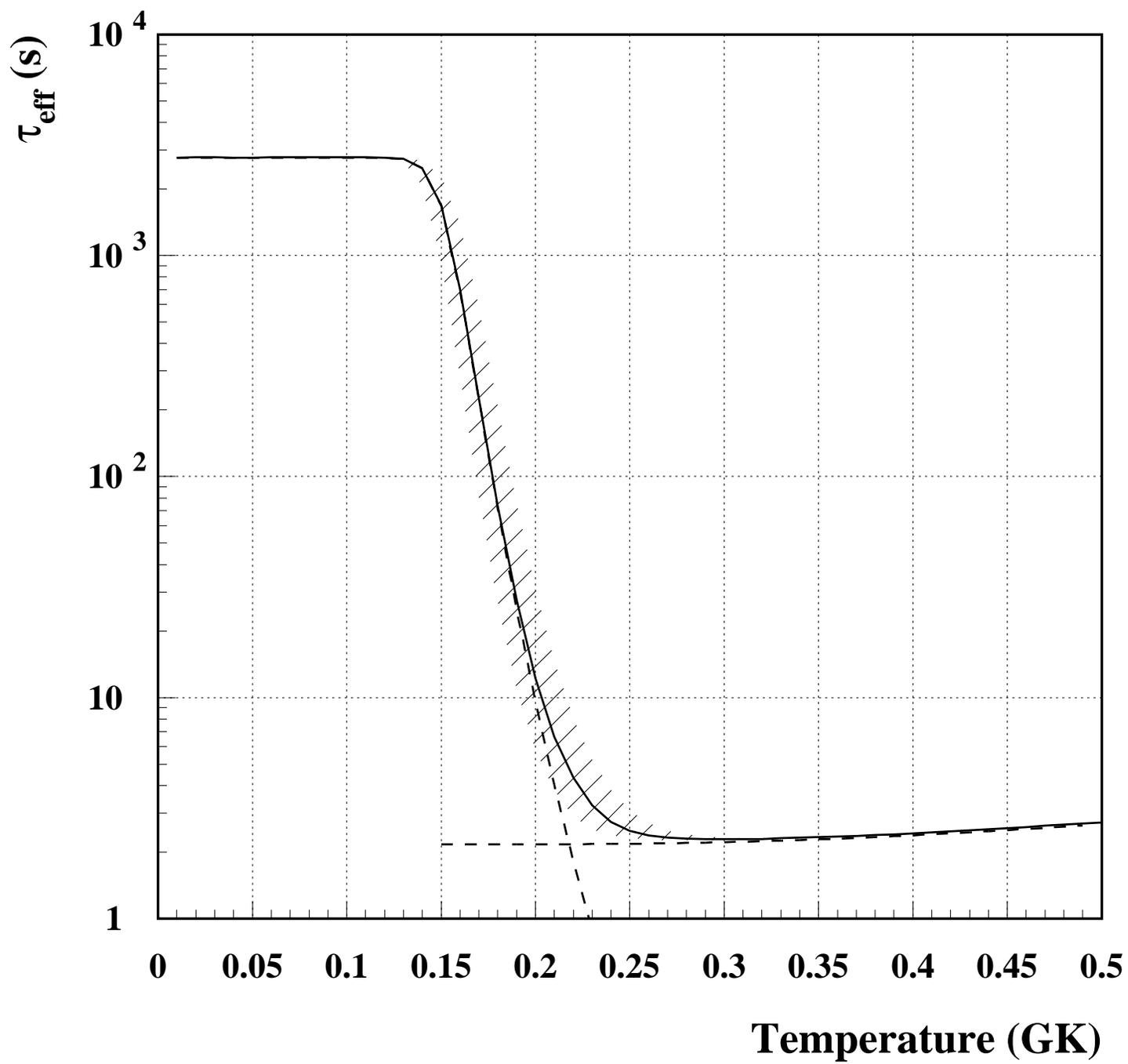
\end{document}